\documentclass[a4paper,11pt]{article}
\usepackage{pos}

\title{Luminosity at LHCb in Run~3}

\author[*]{E. Graverini}

\affiliation{\'{E}cole Polytechnique F\'{e}d\'{e}rale, Lausanne, Switzerland}

\emailAdd{elena.graverini@cern.ch}

\note[*]{Speaker, on behalf of the LHCb Collaboration.\newline{}\indent\hspace*{1.25ex} The author acknowledges the support provided by the Swiss National Science Foundation (SNSF)\newline{}\indent\hspace*{1.25ex} with grant numbers 190734 and 202065.}

\abstract{The LHCb detector optimised its performance in Runs~1 and~2 by stabilising the instantaneous luminosity during a fill. This was achieved by tuning the distance between the two colliding beams according to the measurement of instantaneous luminosity from hardware-based trigger counters. The upgraded LHCb detector operates at fivefold instantaneous luminosity compared to the previous runs, and it has a fully software-based trigger. Consequently, a new approach to the luminosity measurement is adopted. New counters, with particular attention to maximum stability in time, and a new dedicated detector have been introduced for Run~3. Additionally, in order to verify linearity from calibration to data taking conditions, per-fill emittance scans are performed. We present an overview of the newly implemented methods for luminosity measurement, as well as early achievements obtained during the first few weeks of data taking.}

\FullConference{%
  41st International Conference on High Energy physics -- ICHEP2022\\
  6-13 July, 2022\\
  Bologna, Italy
}

\graphicspath{%
{./figs/}
}

\usepackage{xspace}
\renewcommand{\L}{\ensuremath{\mathcal{L}}\xspace}
\newcommand{\muvis}{\ensuremath{\mu_{vis}}\xspace}
\newcommand{\mupvis}{\ensuremath{{\mu'}_{vis}}\xspace}
\newcommand{\svis}{\ensuremath{\sigma_{vis}}\xspace}

\usepackage[backend=biber,
            sorting=none,
            maxnames=50,
            style=numeric-comp,
            sortcites,
            giveninits]{biblatex}
\DeclareFieldFormat[misc,online,proceedings,report]{title}{\mkbibquote{#1}}

\DefineBibliographyStrings{english}{%
  mathesis = {Master's thesis},
}
\renewbibmacro{in:}{%
  \ifentrytype{article}{}{\printtext{\bibstring{in}\intitlepunct}}}

\usepackage{lineno}

\let\oldequation\equation
\let\oldendequation\endequation
\renewenvironment{equation}
  {\linenomathNonumbers\oldequation}
  {\oldendequation\endlinenomath}
\let\oldalign\align
\let\oldendalign\endalign
\renewenvironment{align}
  {\linenomathNonumbers\oldalign}
  {\oldendalign\endlinenomath}

\begin{document}
\maketitle

\section{Luminosity at LHCb}\vspace*{-1.5mm}
The LHCb experiment~\cite{LHCb:2012doh} is a single-arm forward spectrometer specifically designed to study the properties of heavy hadron decays. The experiment tracking system is composed of a silicon pixel vertex detector (VELO), a silicon strip upstream tracker placed before a 4~Tm dipole magnet, and a downstream tracker based on scintillating fiber technology (SciFi). Particle identification is provided by two ring-imaging Cherenkov detectors sensitive in different regions of particle momenta, by an electronic and a hadronic calorimeter, and by a muon detector. The LHCb detector operates at the LHC interaction point IP8.

Together with the beam energy $E_b$, luminosity (\L) is a parameter that quantifies the performance of a particle accelerator: it measures its ability to produce collisions. The rate of process $j$ is related to its cross section $\sigma_j$ by the luminosity:
\begin{align}
R_j = \L \sigma_j.
\end{align}
A precise measurement of this quantity is therefore needed to measure any production cross section.

The LHCb apparatus has a pseudorapidity acceptance of $2\leq\eta\leq 5$, designed for detecting heavy hadrons which are mainly produced in the forward region. To optimize its performance, LHCb operates at constant luminosity throughout a fill, lower with respect to ATLAS and CMS.
To achieve this, a method known as \textit{luminosity levelling} was developed~\cite{Follin:2014nva}. Beams are continuously steered, based on real time feedback from the experiment, in order to achieve a constant luminosity value and thus optimise the LHCb tracking performance.

Luminosity can be measured from the visible number of interactions per bunch crossing \muvis for a given process with effective cross section \svis as
\begin{align}
\L = f \frac{\muvis}{\svis},
\end{align}
where $f$ is the LHC revolution frequency. Here, \muvis can be monitored continuously and used for luminosity levelling. Suitable observables used to measure luminosity are referred to as \textit{luminometers}. An absolute luminosity calibration is achieved by measuring \svis with dedicated procedures. At LHCb, beam-gas imaging has been used in LHC Run~1, and van der Meer scans in Runs~1 and~2~\cite{LHCb:2014vhh}. Both will be employed in Run~3 and beyond.

\begin{figure}\vspace{-3mm}
\centering\includegraphics[width=.73\textwidth]{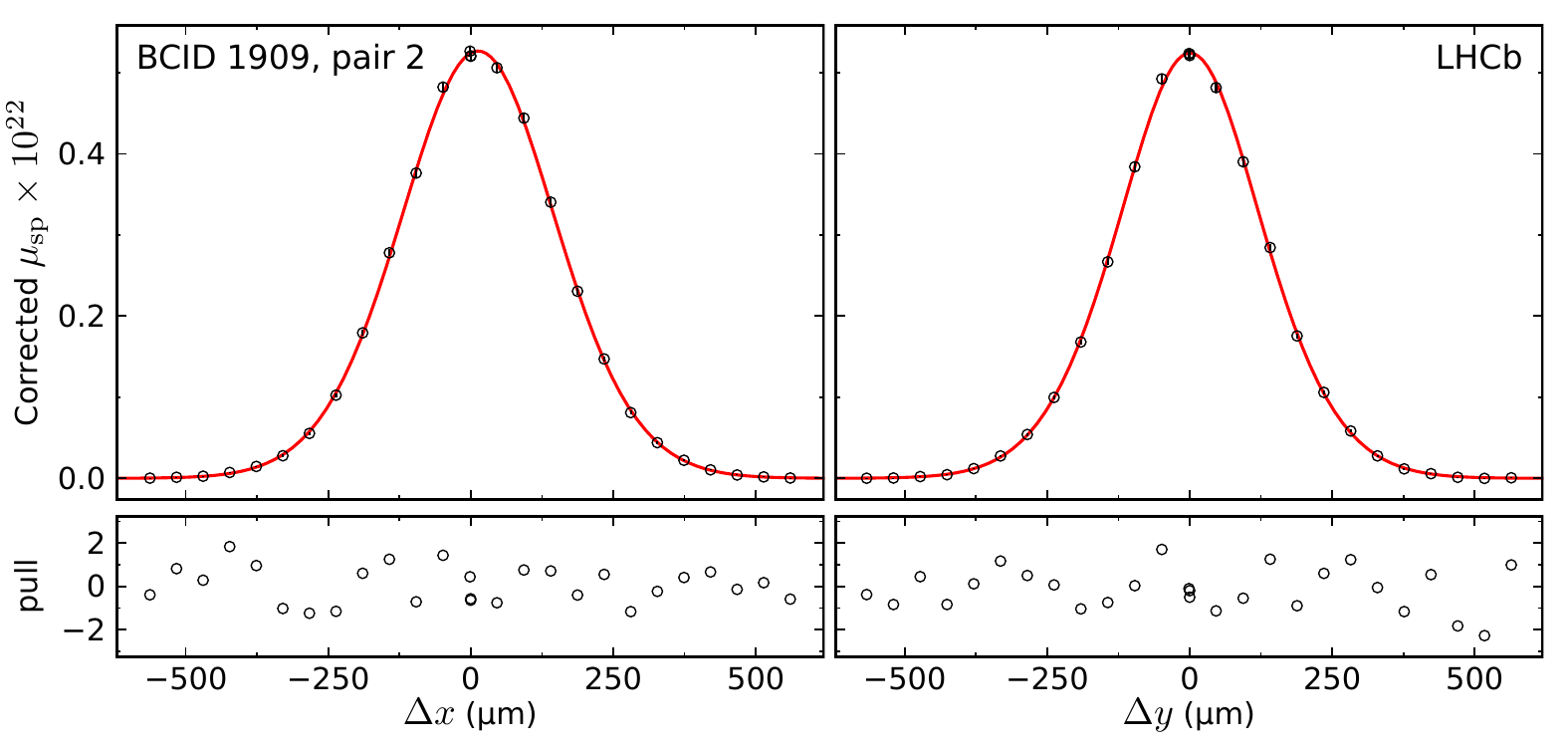}\vspace*{-1.mm}
\caption{VdM profiles along $x$ (left) and $y$ (right) for a single-bunch pair in one scan performed in Run~1~\cite{LHCb:2014vhh}.}\label{fig:1dvdm}\vspace*{-1.mm}
\end{figure}

\subsection{Absolute luminosity calibration}\vspace*{-.5mm}

Van der Meer (VdM) scans are thoroughly described in~\cite{Balagura:2020fuo}. They have been performed once per year and per energy throughout Runs~1 and~2. During these scans, \svis is measured by displacing the two beams transversely with respect to one another in steps, and recording the varying rate of observed interactions: 
\begin{align}
\svis = \frac{\int \mupvis\left(x,y_0\right)dx \int \mupvis\left(x_0,y\right)dy}{\mupvis\left(x_0,y_0\right)}\label{eqn:vdm1d}
\end{align}
where $\mupvis\equiv\muvis/N_1N_2$, with $N_{1,2}$ being the populations of colliding bunches at each step, and the integrals run through the beams displacement along the $x$ and $y$ axes.
Multiple luminometers are calibrated at the same time, therefore \muvis is measured with all available counters, obtaining the respective \svis values.
\figurename~\ref{fig:1dvdm} shows a pair of such scans performed during Run~1, where a double Gaussian beam model is fitted to background-subtracted data points~\cite{LHCb:2014vhh}.
The approach of Eqn.~\ref{eqn:vdm1d} assumes that the $x$ and $y$ profiles are factorizable. However, the systematic uncertainty associated with this assumption was proven to be nonnegligible~\cite{LHCb:2014vhh}. During the LHC Run~2, LHCb has pioneered two-dimensional van der Meer scans~\cite{lumidays2d}, where the beams displacements along $x$ and $y$ are scanned in a 2D grid. This proved to be the most accurate luminosity calibration method. Other LHC experiments will be adopting these 2D scans from Run~3 onwards. 

The increase in luminosity foreseen for Run~3 poses an additional challenge. For LHCb, the average visible number of interactions $\langle\mu\rangle$ increases from $\sim$1 to $\sim$5. Van der Meer scans are usually performed at lower $\mu$, typically $\mu\leq 0.5$; therefore, luminometers must behave linearly with $\mu$ in order to be able to extrapolate their VdM calibrations up to physics $\langle\mu\rangle$. To address this, shorter van der Meer scans, \textit{emittance scans}, will be performed in the same beam conditions used for physics at the beginning or end of each fill.

\begin{figure}\centering
\raisebox{-.5\height}{\includegraphics[height=4.cm]{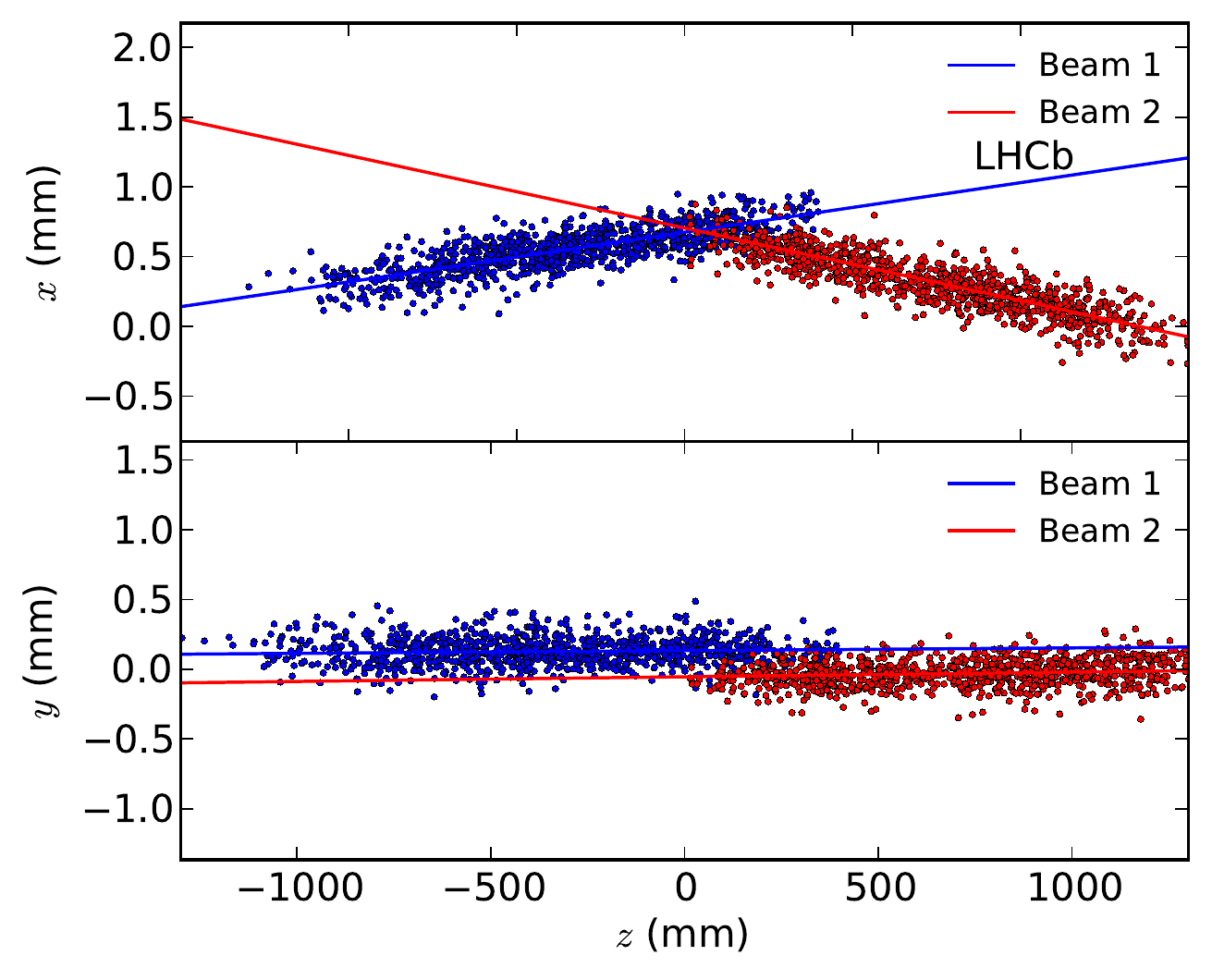}}
\raisebox{-.5\height}{\includegraphics[height=4.cm]{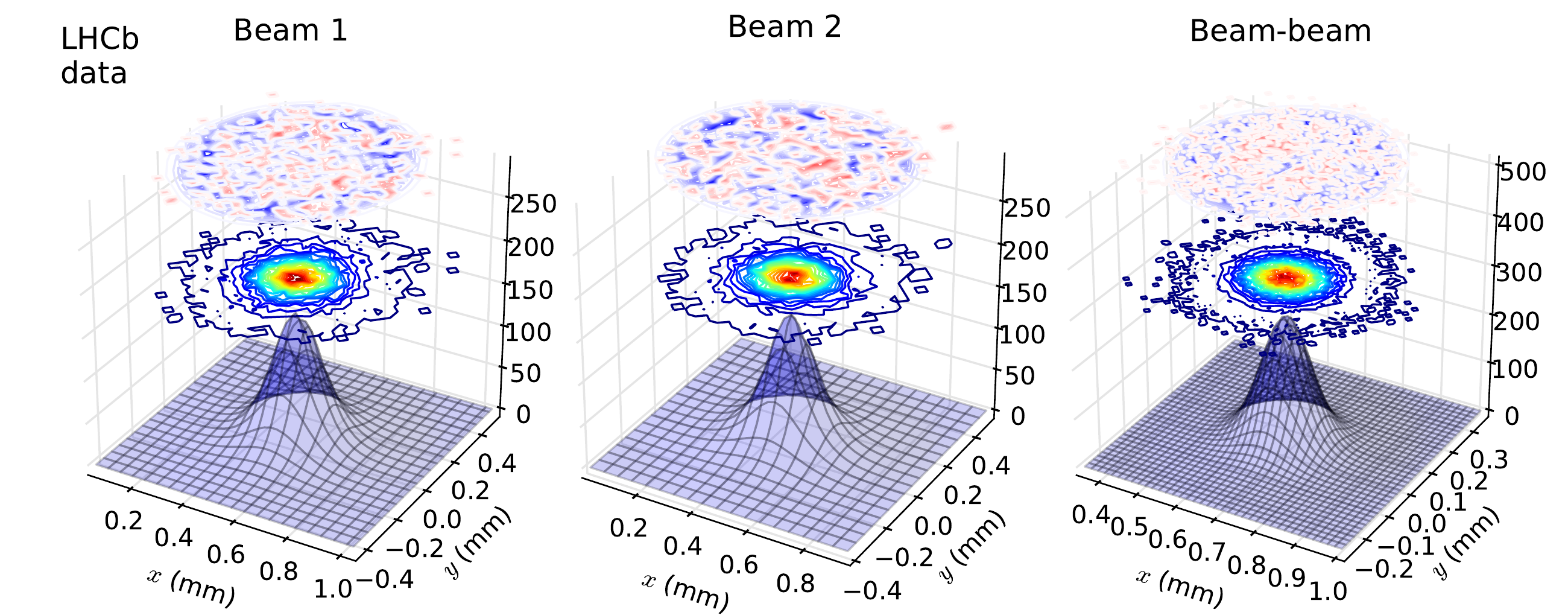}}
\caption{Position of beam-gas vertices in the $xz$ and $yz$ planes during fill 2852 (left). Resolution-corrected fitted shapes of the two beams and of the luminous region, projected onto the central $z$ slice (right)~\cite{LHCb:2014vhh}.}\label{fig:bgi}\vspace*{-2.mm}
\end{figure}

Another approach to calibrate the absolute scale of luminosity, one unique to LHCb, is made possible by the excellent resolution of the LHCb vertex detector. In a technique called \textit{beam-gas imaging} (BGI), interaction vertices between the LHC beam and residual or injected gas molecules are reconstructed~\cite{Barschel:2014iua,LHCb:2014vhh}. \figurename~\ref{fig:bgi} showcases the determination of the shape of the luminous region using BGI. A dedicated system for gas injection, SMOG, was originally conceived to calibrate luminosity with BGI measurements. It has, though, demonstrated great potential for fixed-target physics, pursued by designing the new storage cell SMOG2~\cite{LHCbCollaboration:2673690}, now installed for Run~3. This feature makes LHCb the first experiment that can operate both in collider and in fixed-target mode.

\subsection{Relative luminosity measurement}\vspace*{-.5mm}
For each luminometer adopted during data taking, the measurement of luminosity is given by the measured \svis and by the rate of visible interactions \muvis.
Luminometers must be proportional to \L, and have efficiencies stable in time. A way to mitigate potential nonlinearities and instability is to exploit Poisson statistics and only count the number of empty events. The number of interactions is Poisson distributed around \muvis, and thus 
\begin{align}
P\left(k\vert\muvis\right) = \frac{\muvis^k e^{-\muvis}}{k!} \Longrightarrow P\left(0\vert\muvis\right) = e^{-\muvis} \Longrightarrow \muvis = -\log P\left(0\right),
\end{align}
where $P\left(0\right)$ is the fraction of empty events recorded by the luminometer. This method works very well for low \muvis, but it becomes challenging in Run~3 with $\muvis\!\sim\!5$.

\begin{figure}\vspace*{-3mm}
\centering
\raisebox{-.5\height}{\includegraphics[height=5.25cm]{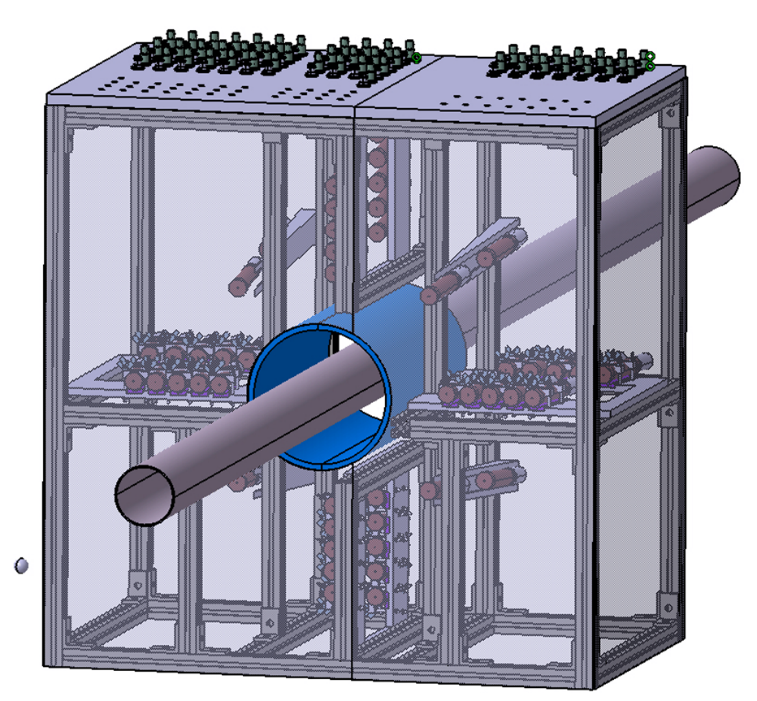}}\hspace*{1cm}
\raisebox{-.5\height}{\includegraphics[height=5.25cm]{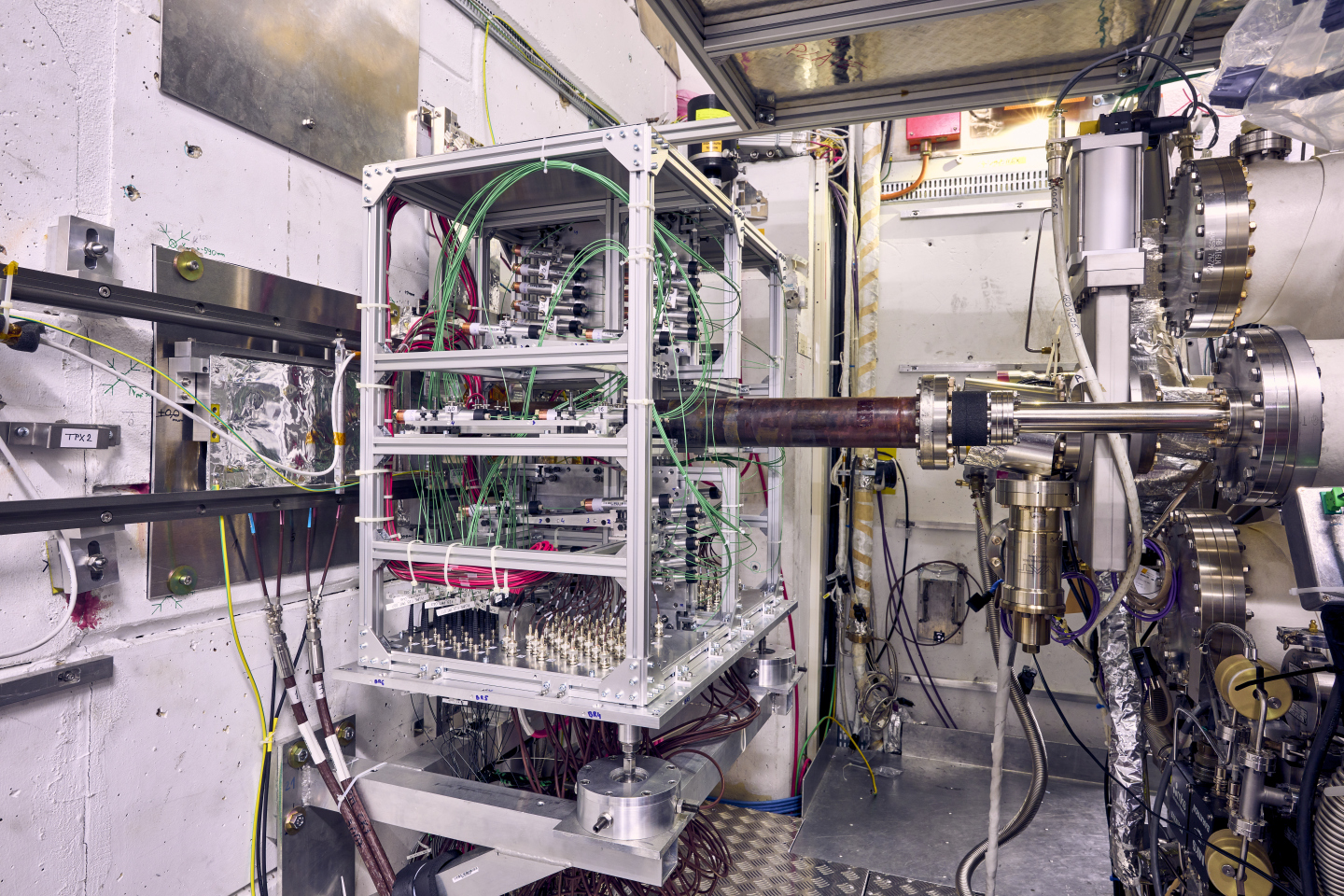}}
\caption{PLUME arrangement (left) and picture of PLUME installed around the beam pipe at IP8 (right).}\label{fig:plume}\vspace*{-2.mm}
\end{figure}

For LHC to perform luminosity levelling at IP8, a precision lower than 5\% is required on real-time \muvis.
During Runs~1 and~2, this measurement was mainly based on the transverse energy deposited in the calorimeters, which was measured at the hardware trigger stage. LHCb moved to a software-only trigger setup in Run~3, completely dropping the hardware-based system in favour of more flexibility for the physics programme~\cite{LHCbCollaboration:2717938}. The data flow has been redesigned: events of interest are now selected for persistence based on offline-quality real time event reconstruction, running on a computing farm comprising both CPUs and GPUs.
Beside the obvious computing challenges, this change implies that a new solution is needed to control luminosity and provide fast, reliable real time feedback to LHC. A dedicated new detector named PLUME (Probe for LUminosity MEasurement), shown in \figurename~\ref{fig:plume}, has been conceived in 2019~\cite{LHCbCollaboration:2750034} and is now installed upstream of the VELO and undergoing commissioning on collision data since May 2022. This detector is in charge of providing real time \muvis measurement for luminosity levelling.

\begin{figure}\vspace*{-3mm}
\centering
\raisebox{-.5\height}{\includegraphics[width=.63\textwidth]{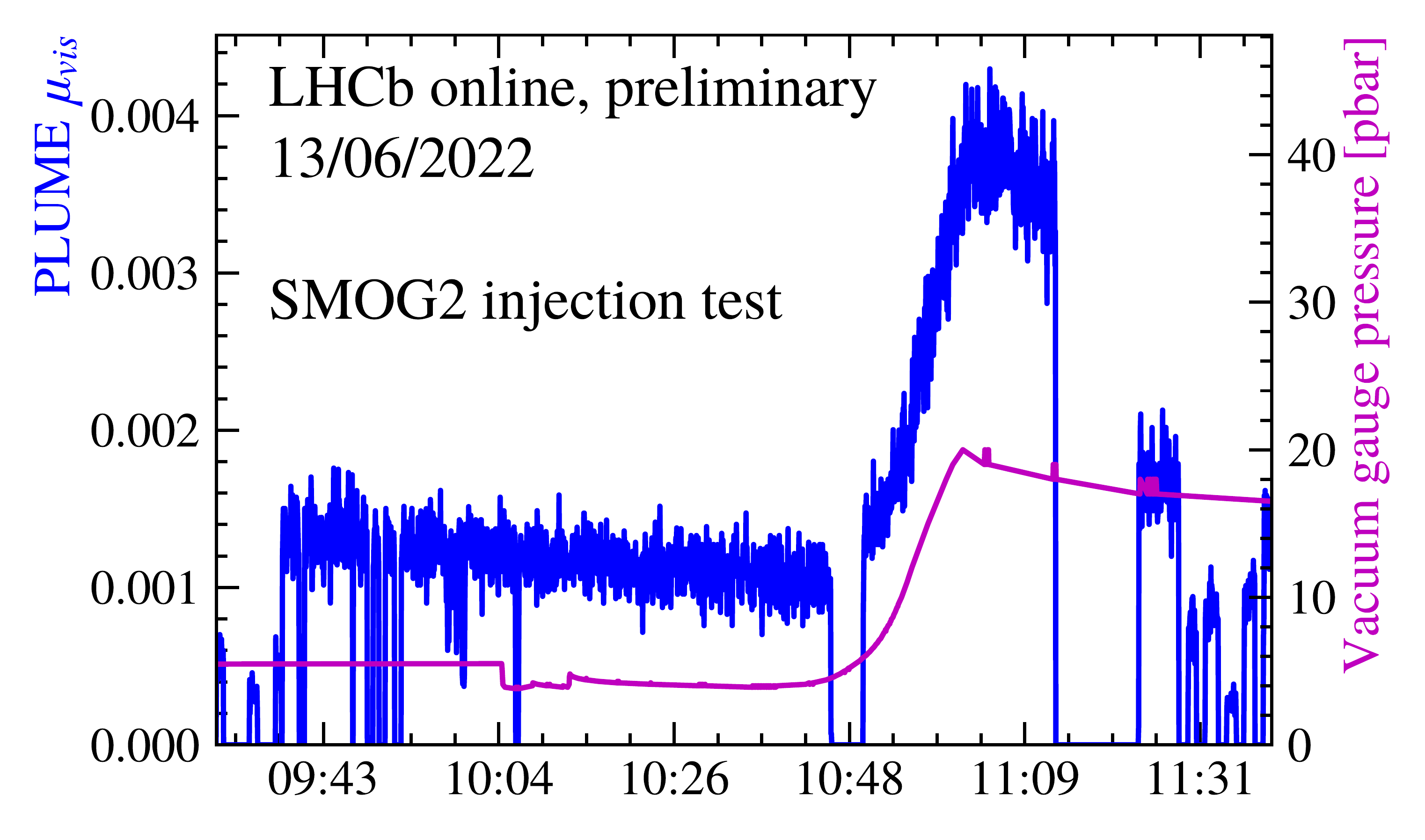}}\hspace*{.5cm}
\raisebox{-.5\height}{\includegraphics[width=.35\textwidth]{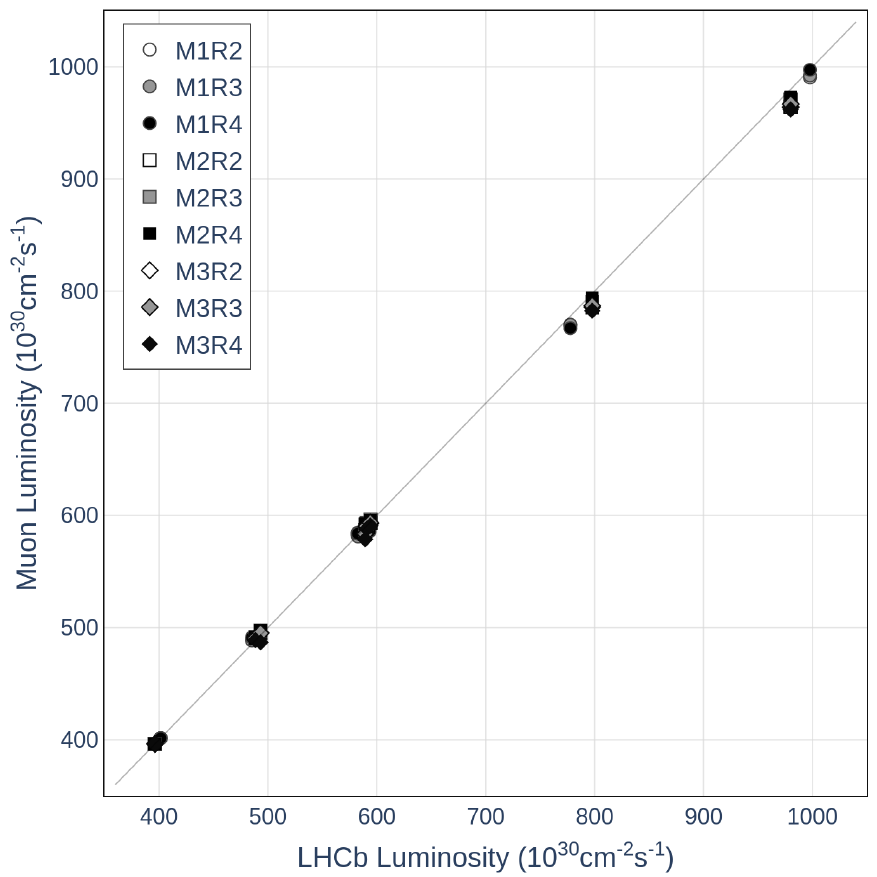}}
\caption{Evolution of gas pressure and rate of interactions recorded in PLUME during a gas injection test with SMOG2 (left). Linearity of a luminometer based on muon hits, from special runs in 2012 (right) \cite{Albicocco:2022gha}.}\label{fig:smog-lumuonmeter}\vspace*{-2.mm}
\end{figure}

PLUME is a hodoscope of 24 projective pairs of Hamamatsu R760 photomultiplier tubes (PMT) with quartz windows. Additional 5~mm thick quartz tablets are placed in front of the PMTs to increase the light yield. The use of quartz, a Cherenkov radiator, and the choice of PMT technology ensures the required fast detector response, with negligible spillover of detected charge between subsequent bunch crossings. At the same time, this design is sufficiently radiation resistant, and it can withstand the estimated fluence of $10^{14}$ neutrons/cm$^2$ expected at its location. The projective PMT pairs layout has been optimized to obtain an occupancy of $\mathcal{O}\left(1\%\right)$ and enable the use of the $\log P\!\left(0\right)$ method. Two pairs are reserved for timing measurements.
The PLUME detector is read out with the same electronics already employed for the LHCb calorimeters. The PMT gain is stabilized by quantifying the response to either backward tracks reconstructed in the VELO, or ultrarelativistic particles recorded in the projective PMT pair counterpart, and with a dedicated LED light-injection system.  \figurename~\ref{fig:smog-lumuonmeter} (left) shows the evolution of gas pressure and rate of interactions recorded in PLUME during a gas injection test with SMOG2.

\begin{figure}
\centering
\hspace{-.5cm}\raisebox{-.5\height}{\includegraphics[height=4.1cm]{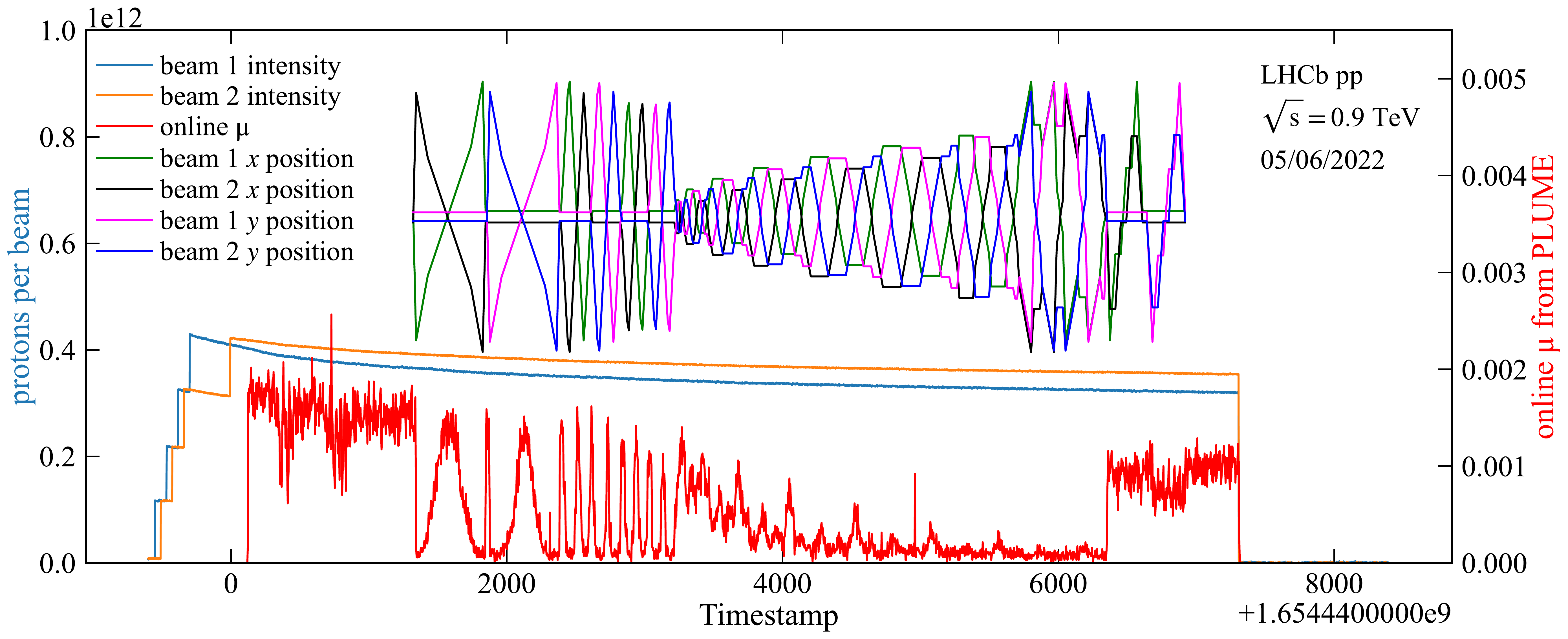}}
\hspace{0cm}\raisebox{-.5\height}{\includegraphics[height=5.1cm]{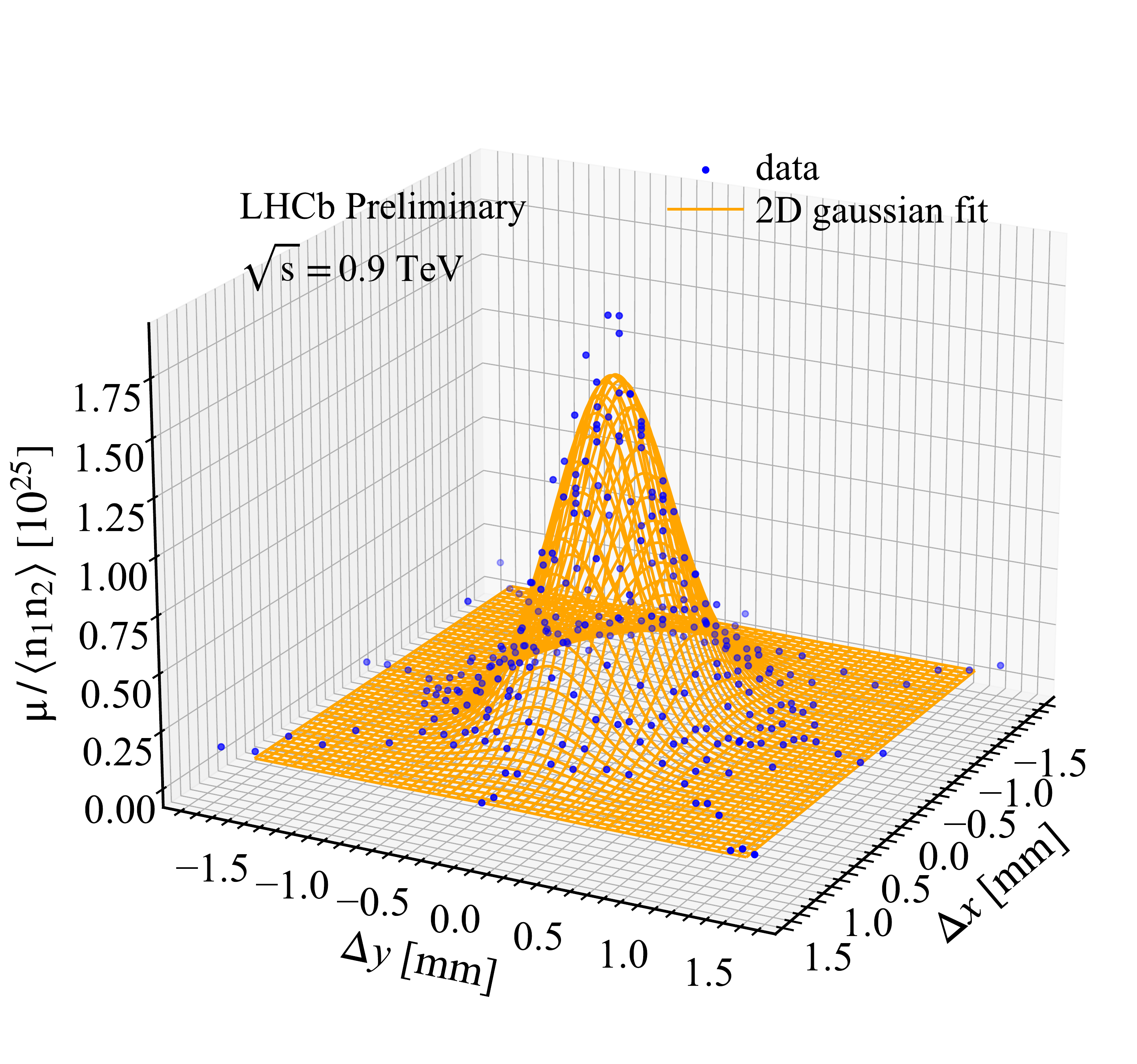}}
\caption{Beam movements and PLUME luminometer value during a VdM session in Run~3 (left). Measured values of $\mu$ normalized to the beam intensities during the 2D VdM scan, fitted to the sum of a constant background and of a double Gaussian signal model (right)~\cite{VY_figure}.}\label{fig:vdm}\vspace*{-2.mm}
\end{figure}

A first calibration of \svis with PLUME was performed on June 5, 2022 using $pp$ collisions at injection energy $\sqrt{s}=900$~GeV. \figurename~\ref{fig:vdm} shows on the same time scale the beams movements and the value of \muvis calculated in real time from the rate of coincidences in 22 PMT pairs. Two 1D van der Meer scans were performed, along the two axes, followed by a longer 2D van der Meer scan and a short lenght scale calibration. The special run lasted about two hours. The analysis of the recorded data is ongoing, but a preliminary fit shows that the values of \svis obtained in the 1D and 2D scans are compatible with each other~\cite{VY_figure}. More VdM scans will be performed at the nominal LHC collision energy $\sqrt{s}=13.6$~TeV and used for physics.

The measurements of production cross sections rely on achieving best possible precision on the integrated luminosity.
In Runs~1 and~2, the luminometers that achieved the best offline precision were the numbers of tracks and vertices reconstructed in the VELO, while many other luminometers have been used for corrections, systematics and cross-checks. An accuracy of 1.16\% was obtained in Run~1~\cite{LHCb:2014vhh}, a record for a bunched hadron collider. The analysis of Run~2 data is being finalized.

In Run~3, the VELO as well as the upstream and downstream trackers have been replaced; the readout electronics of all other subdetectors has been upgraded; full event reconstruction is now performed in real time. An investigation of the performance of many observables proportional to luminosity is ongoing, aimed at adopting them as luminometers. Numbers of tracks and vertices, numbers of coincidences in PLUME, the energy deposited in the calorimeters and the occupancy of various detectors, are under study. A promising study was performed using the muon detector~\cite{Albicocco:2022gha}, whose number of hits exhibits very good linearity as shown in \figurename~\ref{fig:smog-lumuonmeter}.

\section{Conclusions}\vspace*{-1.5mm}

The LHCb experiment has a proven track record for the measurement of luminosity. It has achieved record precision in Run~1 and pioneered two-dimentional van der Meer scans in Run~2. The experiment has made use of unique beam-gas imaging capabilities, and a new storage cell for gas injection has been installed for Run~3. The recent switch to software trigger posed new challenges for real time luminosity monitoring, that have been addressed with the addition of a dedicated detector, PLUME, and the identification of a new set of luminometers. Their properties of linearity and stability are being investigated. A first absolute luminosity measurement has been performed, calibrating PLUME at injection energy.

\vspace*{-1.5mm}
\AtNextBibliography{\small}
\printbibliography[title=References]

@article{Albicocco:2022gha,
    author = "Albicocco, Pietro and others",
    title = "{A Method Based on Muon System to Monitor LHCb Luminosity}",
    doi = "10.3390/sym14050860",
    journal = "Symmetry",
    volume = "14",
    number = "5",
    pages = "860",
    year = "2022"
}

@article{LHCb:2014vhh,
    author = "{LHCb Collaboration}",
    title = "{Precision luminosity measurements at LHCb}",
    eprint = "1410.0149",
    archivePrefix = "arXiv",
    primaryClass = "hep-ex",
    reportNumber = "LHCB-PAPER-2014-047, CERN-PH-EP-2014-221",
    doi = "10.1088/1748-0221/9/12/P12005",
    journal = "JINST",
    volume = "9",
    number = "12",
    pages = "P12005",
    year = "2014"
}

@techreport{LHCbCollaboration:2750034,
      author        = "{LHCb Collaboration}",
      title         = "{LHCb PLUME: Probe for LUminosity MEasurement}",
      institution   = "CERN",
      address       = "Geneva",
      reportNumber  = "CERN-LHCC-2021-002, LHCB-TDR-022",
      year          = "2021",
      doi           = "10.17181/CERN.WLU0.M37F",
}

@techreport{LHCbCollaboration:2673690,
      author        = "{LHCb Collaboration}",
      title         = "{LHCb SMOG Upgrade}",
      institution   = "CERN",
      address       = "Geneva",
      reportNumber  = "CERN-LHCC-2019-005, LHCB-TDR-020",
      year          = "2019",
      doi           = "10.17181/CERN.SAQC.EOWH",
}

@phdthesis{Barschel:2014iua,
    author = "Barschel, Colin",
    title = "{Precision luminosity measurement at LHCb with beam-gas imaging}",
    reportNumber = "CERN-THESIS-2013-301",
    school = "RWTH Aachen U.",
    year = "2014"
}

@unpublished{lumidays2d,
title= {Non factorization in LHCb vdM scans: experience in Run-2, impact on luminosity calibration},
author = "Balagura, Vladislav",
year = {2019},
note= {LHC Lumi Days},
URL= {https://indi.to/VvbF6},
}

@article{Balagura:2020fuo,
    author = "Balagura, Vladislav",
    title = "{Van der Meer scan luminosity measurement and beam\textendash{}beam correction}",
    doi = "10.1140/epjc/s10052-021-08837-y",
    journal = "Eur. Phys. J. C",
    volume = "81",
    number = "1",
    pages = "26",
    year = "2021"
}

@article{LHCb:2012doh,
    author = "{LHCb Collaboration}",
    title = "{Framework TDR for the LHCb Upgrade}",
    reportNumber = "CERN-LHCC-2012-007, LHCb-TDR-12, CERN-LHCC-2012-007, LHCb-TDR-12",
    month = "4",
    year = "2012"
}

@techreport{LHCbCollaboration:2717938,
      author        = "{LHCb Collaboration}",
      title         = "{LHCb Upgrade GPU High Level Trigger Technical Design
                       Report}",
      institution   = "CERN",
      reportNumber  = "CERN-LHCC-2020-006, LHCB-TDR-021",
      address       = "Geneva",
      year          = "2020",
      doi           = "10.17181/CERN.QDVA.5PIR",
}

@misc{VY_figure,
    author = "{LHCb Collaboration}",
    title = "{First absolute calibration of the PLUME online luminometer in Run~3}",
    institution = "CERN",
    url = "https://cds.cern.ch/record/2813388",
    reportNumber = "LHCb-FIGURE-2022-012",
}

@article{Follin:2014nva,
    author = "Follin, F. and Jacquet, D.",
    title = "{Implementation and experience with luminosity levelling with offset beam}",
    eprint = "1410.3667",
    archivePrefix = "arXiv",
    primaryClass = "physics.acc-ph",
    year = "2014"
}
\vspace*{-7.5mm}
\end{document}